# Analysis and implementation of the Buck-Boost Modified Series Forward converter applied to photovoltaic systems

D. López del Moral, A. Barrado, M. Sanz, A. Lázaro, P. Zumel

Universidad Carlos III de Madrid
Electronic Technology Department
Power Electronics Systems Group
Avda. Universidad, 30; 28911, Leganés, Madrid, SPAIN
E-mail: dmoral@ing.uc3m; andres.barrado@uc3m.es

**Abstract**

*The mismatching phenomenon is one of the main issues in photovoltaic (PV) applications. It could reduce the generated power of a string when a PV panel has different performances from the other PV panels connected to the same string. Distributed Maximum Power Point Tracking (DMPPT) architectures are one of the most promising solutions to overcome the drawbacks associated with mismatching phenomena in PV applications. In this kind of architectures, a DC-DC module integrated converter (MIC) manages each PV panel, isolating it from the rest of the PV panels, for harvesting the maximum available power from the Sun. Due to the high number of DC-DC converters used in a grid-tied PV installation, the most desired MIC requirements are high efficiency, low cost and the capability of voltage step-up and step-down.*

*This paper proposes the Buck-Boost Modified Forward (BBMSF) converter as a good candidate to be applied in DMPPT architectures. A complete analysis of the BBMSF converter is carried out, including the steady-state analysis as well as the small signal analysis in continuous conduction mode.*

*The main advantages of the BBMSF converter are its step-up and step-down voltage transfer function; a higher simplicity, since it only includes a single controlled switch; the soft switching characteristics in all the diodes and MOSFET, reaching in some cases ZVS and ZCS, and yielding high efficiencies; the use of an autotransformer, with better performances than a typical Forward transformer; and the good dynamic performance, like the Forward converter ones.*

*The theoretical analyses are validated through the experimental results in a 225W BBMSF prototype designed and built under the requirements of a 100kW grid-tied PV installation, achieving an efficiency up to 93.6%.*

## 1. Introduction

Nowadays the world pays growing attention to the photovoltaic (PV) energy source due to it is clean and practically inexhaustible. Motivated by the increasing tendency of PV installations, that reached up to 229

$GW_{DC}$ at the end of 2016, see Figure 1, interdisciplinary research is continuously developed to sustain the improvement of existing PV conversion technologies and the development of new ones [1].

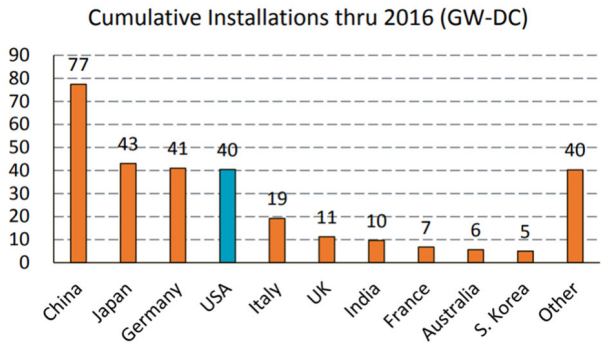

Figure 1. Cumulative PV installations thru 2016 (GW-DC) [1]

One of the most important issues related to PV installations is known as mismatching. The mismatching phenomenon refers to any difference between the PV panels connected to a PV installation.

There are many mismatching causes such temperature differences, dirt, shadows, ageing, etc., [2] - [3]. Any of them changes the PV panel electrical characteristics [4] - [7].

High power PV grid-tied installations use a large number of PV panels. Several PV panels are connected in strings to achieve the desired inverter input voltage. Depending on the inverter voltage and the PV installation power, several strings could be needed. PV panels connected to the same string share the same output current. Therefore, if one of the PV panels connected to the string has different electrical characteristics, it also affects to the rest of the string PV panels, reducing the generated power drastically, even when the irradiation conditions are optimum.

DC-DC converters can be used to isolate PV panels from the others and to make them work at their maximum power point to overcome the mismatching issue. Several PV grid-tied architectures can be considered, depending on the converter integration in the PV installation: the centralised inverter architecture, the converter-per-string architecture and the converter-per-PV panel architecture, also known as Distributed Maximum Power Point Tracking (DMPPT) architecture [8] - [9]. The deeper the converter integration, the better the available power is harvested regardless of the mismatching issue. Therefore, the DMPPT architecture, see Figure 2, is the one that completely isolates each PV panel from the others, allowing for harvesting the maximum available power [10] - [14]. Nevertheless, one DC-DC converter per PV panel or module integrated converter (MIC) is required, increasing the cost of the installation.

The desired MIC requirements are high efficiency, low cost and the capability of voltage step-up and step-down.

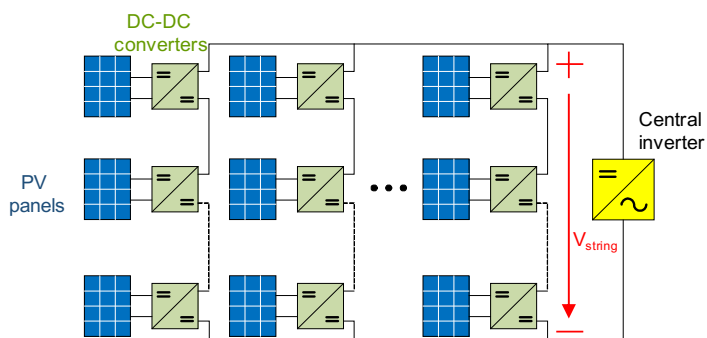

Figure 2. PV grid-tied installation with the DMPPT architecture

Several authors have focused their research on high-efficiency converters. Some of these high-efficiency topologies are based in Partial-Power Conversion (PPC) converters, also referred as Series Connection converters, Parallel-Power-Processed (PPP) converters or Direct Energy Transfer (DET) converters [15] – [23].

In all of them, the converter only manages a part of the energy improving the efficiency, whereas the rest of the output power is directly delivered to the load. The best efficiency achieved in these converters is up to 98% in [16]. The main constraint of based PPC topologies is that it can only be used to voltage step-up, reducing the string-configuration possibilities in the PV installation.

Other authors have also obtained efficiencies around 98% with full power processing topologies, but these topologies are only capable of voltage step-up or step-down [24] - [26].

However, although any DC-DC converter can be employed in PV installations, the highest flexibility regarding the number of PV panels per string is only achieved with voltage step-up and step-down converters [27] - [30]. Within the voltage step-up and step-down topologies, one of the most promising ones is the classical Non-Inverting Buck-Boost converter [27] and [31]. Although very high efficiencies have been obtained in [27], some drawbacks of this topology are the high current that flows through the inductor and switches, and that it requires four switches and drivers. Therefore, the complexity and the components count increases.

In this paper, the Buck-Boost Modified Forward (BBMSF) converter is proposed as a good candidate to be applied in DMPPT architectures [32], after considering this state-of-the-art. It is a high-efficiency DC-DC converter capable of both, voltage step-up and step-down. As an advantage, it only has one active switch, so it only requires one driver. Also, thanks to the autotransformer connection, only a part of the output power is magnetically processed, reducing the power losses and improving the autotransformer performances.

A complete analysis of the BBMSF converter is carried out in Section 2, including the time domain analysis as well as the frequency domain analysis in continuous conduction mode. The theoretical analysis is verified in Section 3, through a 225W BBMSF prototype designed and built under the requirements of a 100kW grid-tied PV installation, considering the effect of different mismatching ratios. Finally, in Section 4, the main conclusions of the research are summarised.

## 2. THEORETICAL ANALYSIS OF THE BBMSF CONVERTER

This paper introduces the Buck-Boost Modified Series Forward (BBMSF) converter. The main difference between the Forward converter and this topology is the use and connection of the autotransformer, instead of the classic transformer, see Figure 3. Thanks to this connection, a part of the power delivered to the load is not magnetically processed. Therefore, the size of the autotransformer is lower than the classic Forward converter one. Also, the losses are reduced.

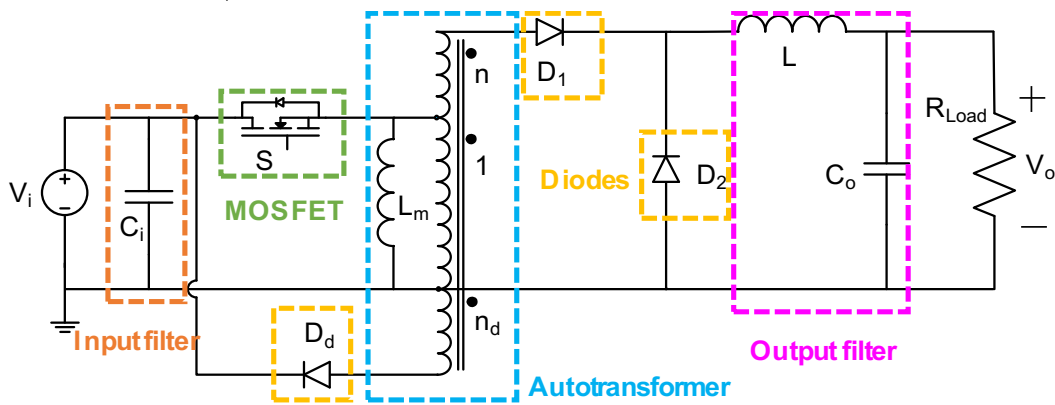

Figure 3. BBMSF electrical scheme

As explained in section 2.2, the input-output transfer function of the BBMSF converter is like the Forward one, with the autotransformer contribution. Therefore, it keeps the advantage of having a voltage step-up and step-down performance. On the other hand, the autotransformer does not isolate source and load. It can be seen in Figure 3 that the input and output voltages are referred to the same ground connection. Another difference between the Forward topology and the BBMSF topology is the location of the MOSFET. For the proper operation of the BBMSF converter, the MOSFET must be placed in the current path between the source and load. Otherwise, the magnetising inductance $L_m$ does not reset properly.

The input and output filters and the diodes are like the Forward converter ones.

## 2.1 PRINCIPLE OF OPERATION

This section analyses each one of the intervals that define the principle of operation of the BBMSF converter in CCM. Two main intervals can be defined depending on the MOSFET state: $t_{ON}$, while the S MOSFET is switched on; and $t_{OFF}$, while it is switched off. Going further in the analysis, the $t_{OFF}$ interval can also be divided into $t_{OFF1}$ and $t_{OFF2}$. These $t_{OFF}$ intervals are related to a change in the magnetising inductance ($L_m$) state. In Figure 4 to Figure 6 are depicted the current paths follow by the current in each interval.

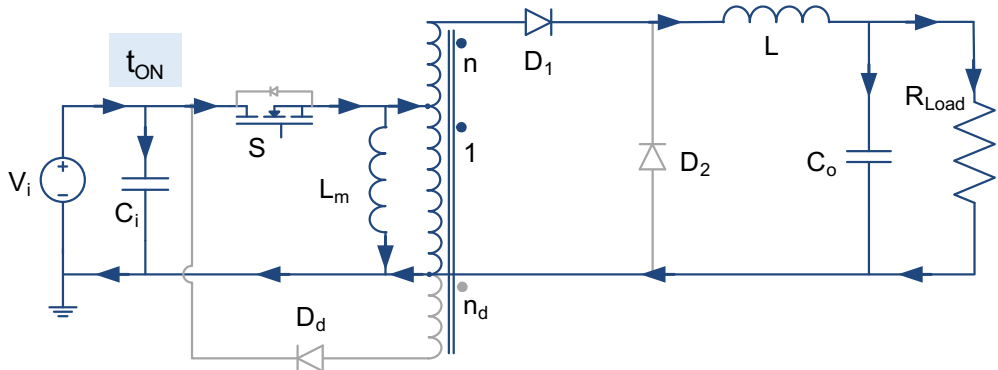

Figure 4. BBMSF current paths during the $t_{ON}$ interval

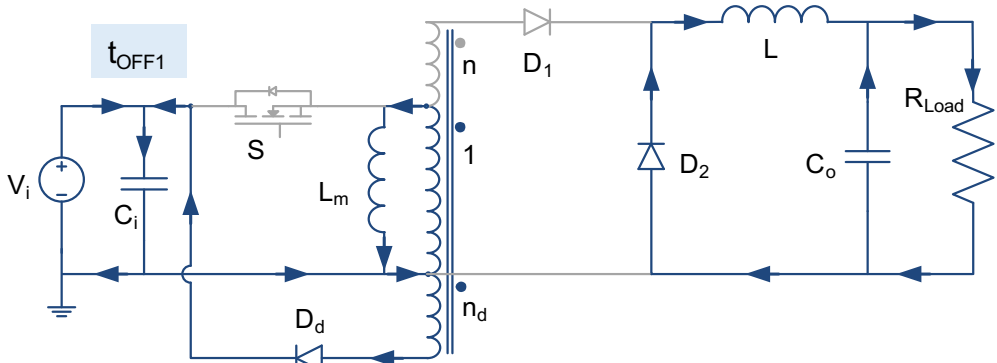

Figure 5. BBMSF current paths during the $t_{OFF1}$ interval

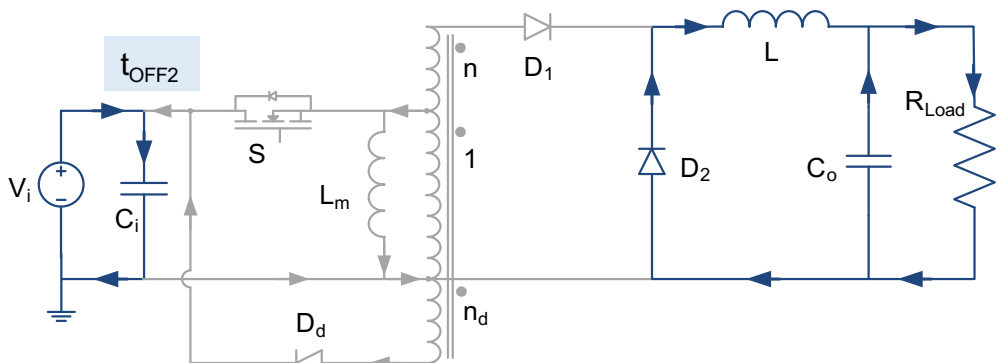

Figure 6. BBMSF current paths during the $t_{OFF2}$ interval

As it can be seen, the current paths are highlighted in blue colour, whereas the rest of the circuit is depicted in grey colour. The direction of the current is denoted with arrows. For the sake of simplicity, the model described in this analysis is ideal, without taking into account the parasitic inductances, capacitances and resistances that are in real converters.

- $t_{ON}$

As it has been mentioned before, the first interval corresponds to the time while the S MOSFET is switched on. During this interval, both inductances, L and $L_m$, are storing energy.

The converter delivers the power from the input to the output filter through the S MOSFET, the autotransformer and the $D_1$ diode, see Figure 4. Thanks to the autotransformer connection, only a part of the output power is magnetically processed, increasing the converter's efficiency and reducing the size of the autotransformer, compared with the Forward converter transformer.

- $t_{OFF1}$

Once the MOSFET is switched off, the reset of the output inductance L and the magnetising inductance $L_m$ begins. As it can be seen in Figure 5, the previously stored energy in $L_m$ is delivered to the input filter through the reset diode $D_d$, after being magnetically processed.

Due to the $D_2$ diode is forward biased, the L current can freely flow, and the stored energy in L is delivered to the output.

- $t_{OFF2}$

The last interval begins when the $L_m$ reset ends. At that moment just the current through the output filter remains flowing, see Figure 6. This interval ends when the S MOSFET switches on again, starting a new cycle.

Table I summarises the main events in each switching interval.

Table I. Summary of the principle of operation events for the BBMSF converter

| Switching interval | Start event | Main considerations | Final event |
|---|---|---|---|
| $t_{ON}$ | S is turned on | L and $L_m$ → store energy<br>Part of the output power → magnetically processed<br>$D_1$ → positive biased | S is turned off |
| $t_{OFF1}$ | S is turned off | L and $L_m$ → deliver energy<br>$D_2$ and $D_d$ → positive biased | $I_{Lm} = 0\ A$ |
| $t_{OFF2}$ | $I_{Lm} = 0\ A$ | $I_{Lm} = 0\ A$<br>L → continue delivering energy<br>$D_2$ → positive biased | S is turned on |

## 2.2 STEADY-STATE OPERATION IN CONTINUOUS CONDUCTION MODE

The main expression that defines a DC-DC converter, while using it as a voltage source, is the output-input voltage transfer function. A voltage-per-second balance in the output filter inductor L is carried out to obtain this expression, see (1).

$$[(1 + n) \cdot V_i - V_o] \cdot D \cdot T = V_o \cdot (1 - D) \cdot T \tag{1}$$

In (2), the BBMSF converter output-input voltage transfer function in Continuous Conduction Mode (CCM) is calculated.

$$\frac{V_o}{V_i} = (1+n) \cdot d \qquad (2)$$

As it can be seen, the output-input voltage transfer function is like the Forward converter one but considering the autotransformer connection, see the $(1+n)$ factor.

Current and voltage stresses are analysed for each one of the BBMSF converter components hereafter.

- Output filter inductance (L)

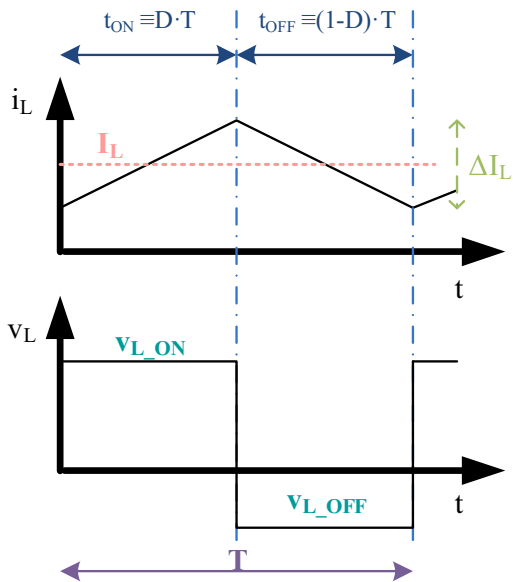

$$\Delta I_L = \frac{V_i \cdot (1+n) \cdot (1-D) \cdot D}{L \cdot f_{sw}} \qquad (3)$$

$$I_L = I_{string} = \frac{P}{V_i \cdot (1+n) \cdot D} \qquad (4)$$

$$v_{L\_ON} = V_i - V_o \qquad (5)$$

$$v_{L\_OFF} = -V_o \qquad (6)$$

Figure 7. Output inductor current and voltage, $i_L$ and $v_L$ respectively, waveforms during one switching period.

- Autotransformer

The design of the autotransformer is crucial. For ensuring that the converter can correctly develop the autotransformer reset, the condition (7) must be fulfilled, regardless of the duty cycle value.

$$n_d \leq \frac{(1-d)}{d} \qquad (7)$$

Several currents can be used to define the autotransformer behaviour. However, only the magnetising voltage and current are obtained in this section. The rest of the autotransformer currents can be obtained from other components: the secondary current is the same as the $D_1$ diode current, the primary current is the same as the secondary current divided by the turns ratio (n), and the tertiary current corresponds with the $D_d$ diode.

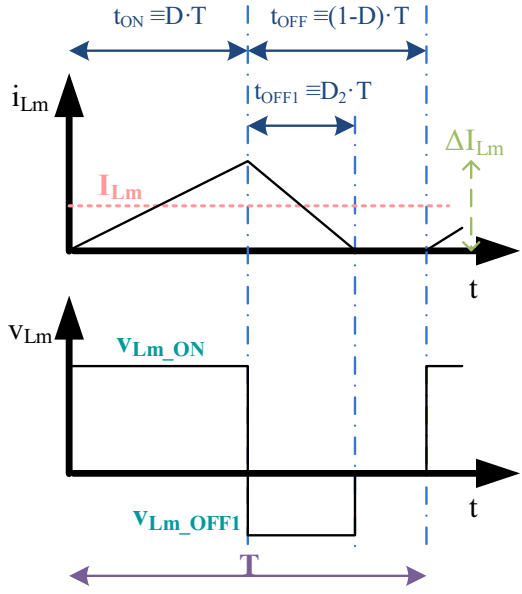

Figure 8. Magnetizing inductor current and voltage, $i_{Lm}$ and $v_{Lm}$ respectively, waveforms during one switching period.

$$\Delta I_{Lm} = \frac{V_i \cdot D}{L_m \cdot f_{sw}} \quad (8)$$

$$I_{Lm} = \frac{\Delta I_{Lm} \cdot (D + D_2)}{V_i} \quad (9)$$

$$D_2 = \frac{n_d}{1 + n_d} \cdot D \quad (10)$$

$$v_{Lm\_ON} = V_i \quad (11)$$

$$v_{Lm\_OFF1} = \frac{V_i}{n_d} \quad (12)$$

- $D_1$ diode

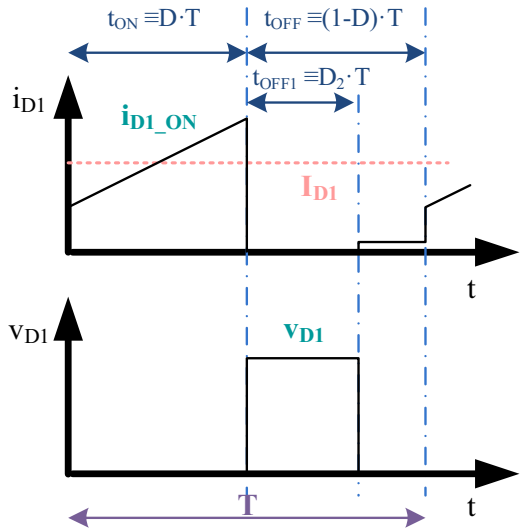

Figure 9. $D_1$ diode current and voltage, $i_{D1}$ and $v_{D1}$ respectively, waveforms during one switching period.

$$i_{D1\_ON} = i_L \quad (13)$$

$$I_{D1} = I_L \cdot D \quad (14)$$

$$v_{D1} = V_i \cdot \frac{(1 + n)}{n_d} \quad (15)$$

- $D_2$ diode

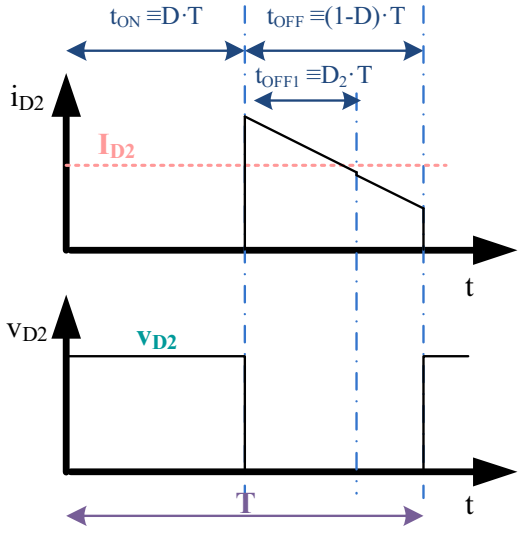

Figure 10. $D_2$ diode current and voltage, $i_{D2}$ and $v_{D2}$ respectively, waveforms during one switching period.

$$i_{D2\_OFF} = i_L \qquad (16)$$

$$I_{D2} = I_L \cdot (1 - D) \qquad (17)$$

$$v_{D2} = V_i \cdot (1 + n) \qquad (18)$$

- $D_d$ diode

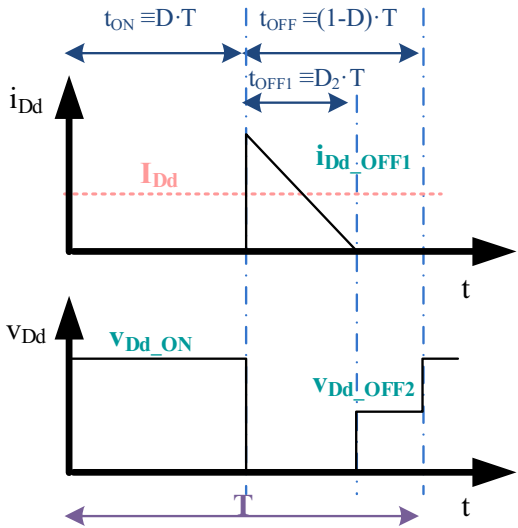

Figure 11. $D_d$ diode current and voltage, $i_{Dd}$ and $v_{Dd}$ respectively, waveforms during one switching period.

$$i_{Dd\_OFF1} = \frac{i_{Lm}}{n_d} \qquad (19)$$

$$I_{Dd} = \frac{\Delta i_{Lm}}{n_d} \cdot D_2 \qquad (20)$$

$$v_{Dd\_ON} = V_i \cdot (1 + n_d) \qquad (21)$$

$$v_{Dd\_OFF2} = V_i \qquad (22)$$

- MOSFET (S)

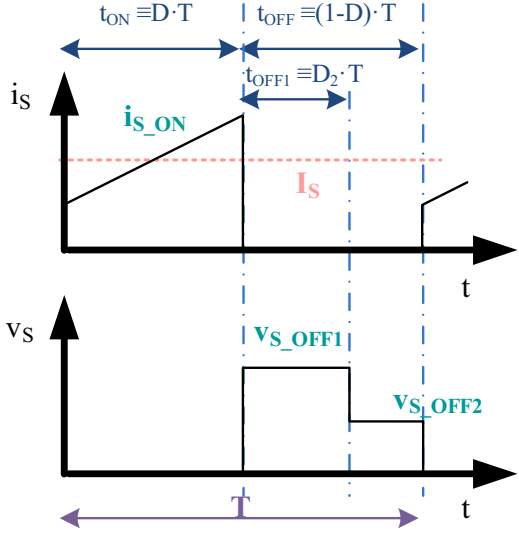

Figure 12. S MOSFET current and voltage, $i_S$ and $v_S$ respectively, waveforms during one switching period.

$$i_{S\_ON} = i_L \cdot (1 + n) + i_{Lm} \qquad (23)$$

$$I_S = I_L \cdot (1 + n) \cdot D + I_{Lm} \qquad (24)$$

$$v_{S\_OFF1} = V_i \cdot \frac{(1 + n_d)}{n_d} \qquad (25)$$

$$v_{S\_OFF2} = V_i \qquad (26)$$

A. Power transfer analysis

The connection of the autotransformer is one of the key points of the BBMSF converter. Thanks to this connection, only a part of the delivered power is magnetically processed by the autotransformer. The size of the autotransformer and its power losses are therefore reduced.

In this section, the theoretical expression that defines the power percentage managed by the autotransformer in CCM is calculated.

Using the equation (2) in the output power expression, (27) is obtained.

$$P_o = V_o \cdot I_o = (1 + n) \cdot D \cdot V_i \cdot I_o = D \cdot V_i \cdot I_o + n \cdot D \cdot V_i \cdot I_o = P_{not\_mag} + P_{mag} \qquad (27)$$

From the equation (27), the relation between the magnetically processed and not magnetically processed power transference, $P_{mag}$ and $P_{not\_mag}$ respectively, is calculated:

$$\frac{P_{mag}}{P_{not\_mag}} = n \qquad (28)$$

The ratio between the magnetically processed and not magnetically processed power for the BBMSF converter topology only depends on the turns ratio $n$. It means that for a fixed turns ratio n, the autotransformer will always manage the same power percentage, regardless of the output voltage conditions. The percentages of both $P_{not\_mag}$ and $P_{mag}$ with respect the output power are described in equations (29) and (30) respectively.

$$P_{not\_mag} = \frac{1}{1 + n} \cdot P_o \qquad (29) \qquad P_{mag} = \frac{n}{1 + n} \cdot P_o \qquad (30)$$

In Table II, several cases regarding turns ratio parameter are selected to illustrate the equations (29) and (30).

Table II. Not magnetically and magnetically processed power percentage

| n | 0.1 | 0.5 | 1 | 1.5 | 2 |
|---|---|---|---|---|---|
| $P_{not\_mag}$ (%) | 0.909 | 0.667 | 0.500 | 0.400 | 0.333 |
| $P_{mag}$ (%) | 0.091 | 0.333 | 0.500 | 0.600 | 0.667 |

As it can be seen, the lower the turns ratio $n$, the lower $P_{mag}$ percentage is. Consequently, for a higher efficiency, lower $n$ values are desired. On the other hand, as it can be deduced from (2), for the same duty cycle value, low turns ratio values reduce the voltage step-up.

Low reset turns ratio $n_d$ must be employed to ensure the autotransformer reset. Due to the dependence between the voltage of some components and the reset turns ratio, very low values of $n_d$ cannot be selected in order to avoid very high voltage stresses in those components, see (15) and (25). To achieve an optimal design, a tradeoff between the not magnetically processed power transference, the highest achievable output voltage and the voltage stresses in the converter components must be found.

## 2.3 SMALL-SIGNAL MODEL. FREQUENCY DOMAIN ANALYSIS

Classic model techniques based on averaging, linearization and perturbation, have been employed to obtain the small signal model of the BBMSF converter in CCM [33] - [35], see Figure 13.

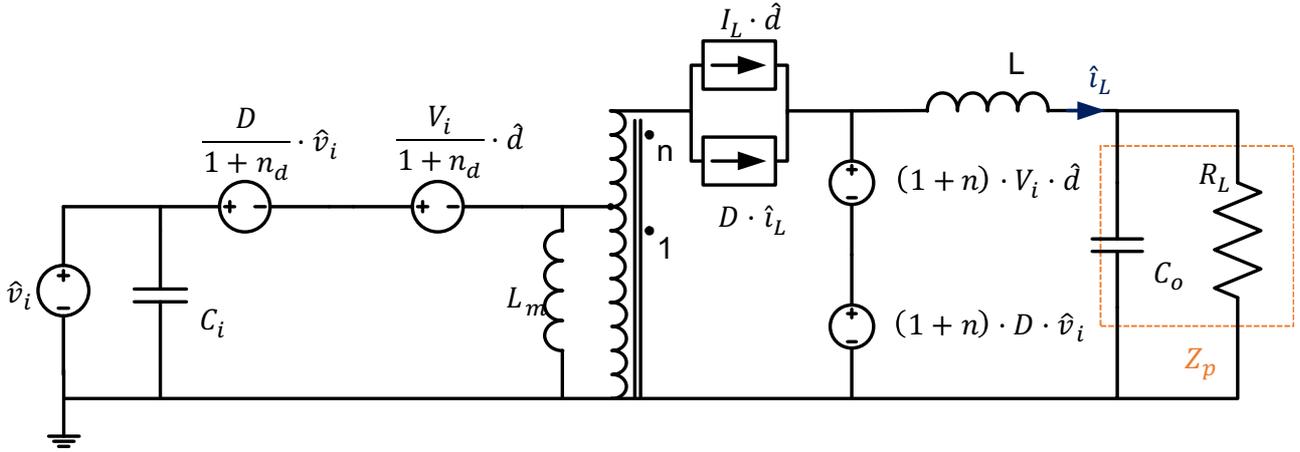

Figure 13. Small signal model of the BBMSF converter

From the small signal model of the BBMSF converter, the expression (31) can be obtained. Note that capital letters refer to DC values for a specific operational point, whereas the disturbed variables are denoted with the superscript "∧". The variable $Z_p$ is defined as the parallel association between the output capacitor $C_o$ and the load resistor $R_L$.

$$\hat{\imath}_L(s) = \frac{(1+n) \cdot V_i}{Z_L(s)} \cdot \hat{d} - \frac{1}{Z_L(s)} \cdot \hat{v}_o + \frac{(1+n) \cdot D}{Z_L(s)} \cdot \hat{v}_i \tag{31}$$

Where $Z_L(s) = s \cdot L$ and $Z_p(s) = \frac{R_L}{1+s \cdot C_o \cdot R_L}$.

As it can be seen in Figure 13, $\hat{\imath}_L(s)$ represents the disturbed current flowing through the load, i.e., the injected current. Once $\hat{\imath}_L(s)$ is defined, the small signal block diagram is depicted to carry on the analysis of the effect of each perturbation separately, see Figure 14.

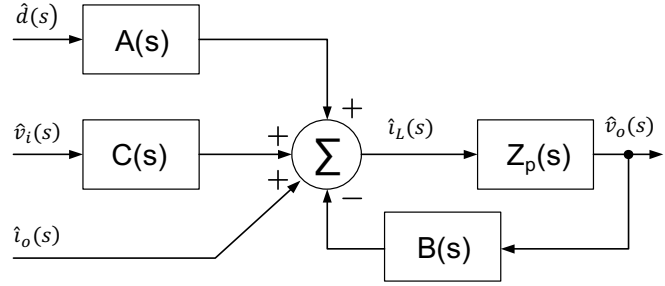

Figure 14. Small-signal block diagram

Paying attention to both the expression (31) and the Figure 14 the expressions of each one of the small signal blocks can be obtained.

Table III. Expressions of the BBMSF converter small-signal blocks

| Small signal block | Expression |
|---|---|
| $A(s)$ | $\dfrac{(1+n) \cdot V_i}{Z_L(s)}$ |
| $B(s)$ | $\dfrac{1}{Z_L(s)}$ |
| $C(s)$ | $\dfrac{(1+n) \cdot D}{Z_L(s)}$ |
| $Z_p(s)$ | $\dfrac{R_L}{1 + s \cdot C_o \cdot R_L}$ |

The most common expressions used for defining the dynamic performances of a converter are the output voltage-duty cycle gain $G_{vd}(s)$, the output-input voltage gain (audiosusceptibility) $G_{vv}(s)$ and the output impedance $Z_o(s)$. These expressions, calculated for the BBMSF converter in CCM, are shown respectively in (32) - (34).

$$G_{vd}(s) = \frac{\hat{v}_o}{\hat{d}} = \frac{A(s) \cdot Z_p(s)}{1 + B(s) \cdot Z_p(s)} = (1+n) \cdot V_i \cdot \frac{w_o^2}{s^2 + s \cdot \frac{1}{R_L \cdot C_o} + w_o^2} \quad (32)$$

$$G_{vv}(s) = \frac{\hat{v}_o}{\hat{v}_i} = \frac{C(s) \cdot Z_p(s)}{1 + B(s) \cdot Z_p(s)} = (1+n) \cdot D \cdot \frac{w_o^2}{s^2 + s \cdot \frac{1}{R_L \cdot C_o} + w_o^2} \quad (33)$$

$$Z_o(s) = \frac{\hat{v}_o}{\hat{\imath}_o} = \frac{Z_p(s)}{1 + B(s) \cdot Z_p(s)} = \frac{1}{C_o} \cdot \frac{s}{s^2 + s \cdot \frac{1}{R_L \cdot C_o} + w_o^2} \quad (34)$$

Where the natural resonant frequency of the system is $w_o = \sqrt{\dfrac{1}{L \cdot C_o}}$.

It is noteworthy that the BBMSF converter $G_{vd}(s)$ expression is like the Forward converter one, see (32). The only difference is the $(1+n)$ factor, which comes from the autotransformer connection.

Some simulations are carried out in next section to validate the theoretical analysis.

### 2.3.1 Simulation validation

Each one of the small signal transfer function expressions, obtained in the previous section in (32) - (34), is depicted in Mathcad® and compared with the simulation results obtained with PSIM®. Table IV summarizes the specific values employed for this comparison.

Table IV. Parameters of the BBMSF converter for a non-shaded PV panel in the Scenario 1

| Parameter | Definition | Value |
| --- | --- | --- |
| $f_{sw}$ | Switching frequency | 50 kHz |
| $V_i$ | Input voltage | 29.3V |
| D | Duty cycle | 0.689 |
| n | Autotransformer secondary turns ratio | 1 |
| L | Output filter inductance | 68 $\mu$H |
| $C_o$ | Output filter capacitance | 112 $\mu$F |
| $R_L$ | Output load | 7.255 Ω |

All the values shown in Table IV corresponds to the ones selected in the prototyping section, for a non-shaded PV panel in the Scenario 1. The case of study and converter specifications are defined in the converters design section.

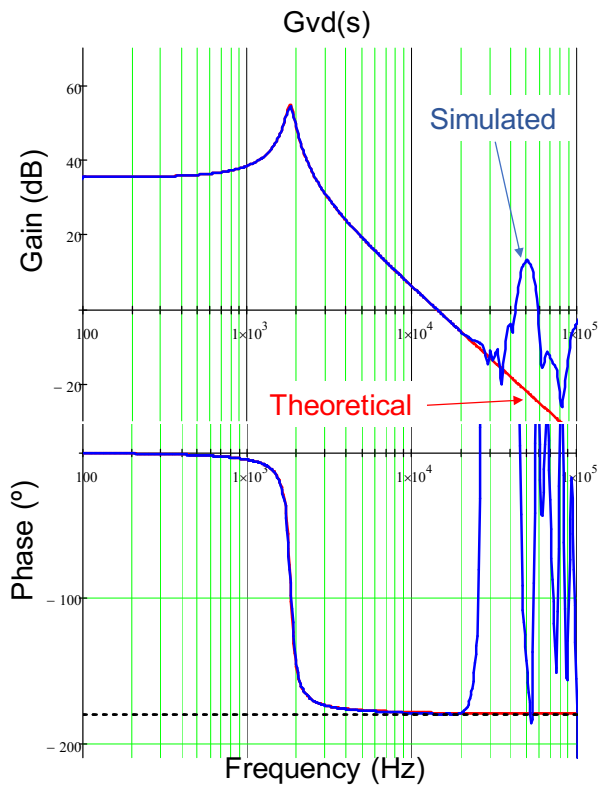

Figure 15. BBMSF converter output voltage-duty cycle small signal transfer function, Gvd(s)

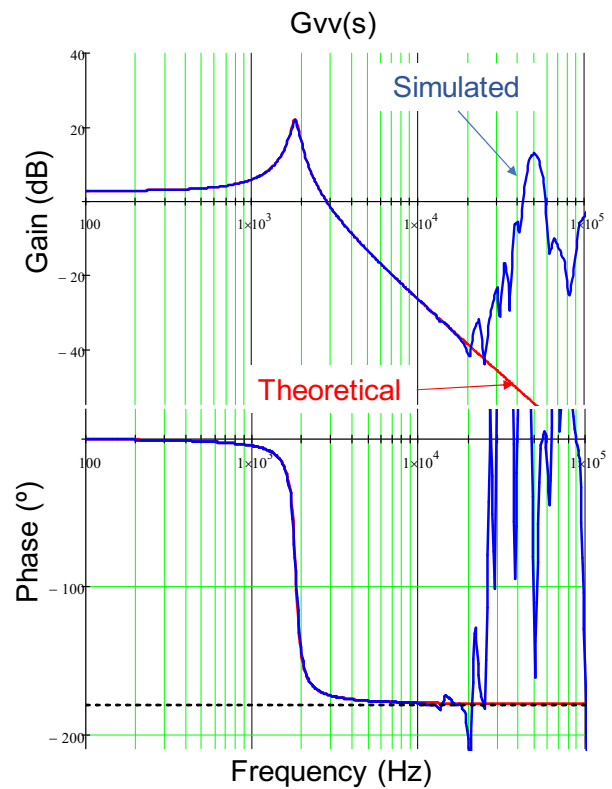

Figure 16. BBMSF converter audiosusceptibility, Gvv(s)

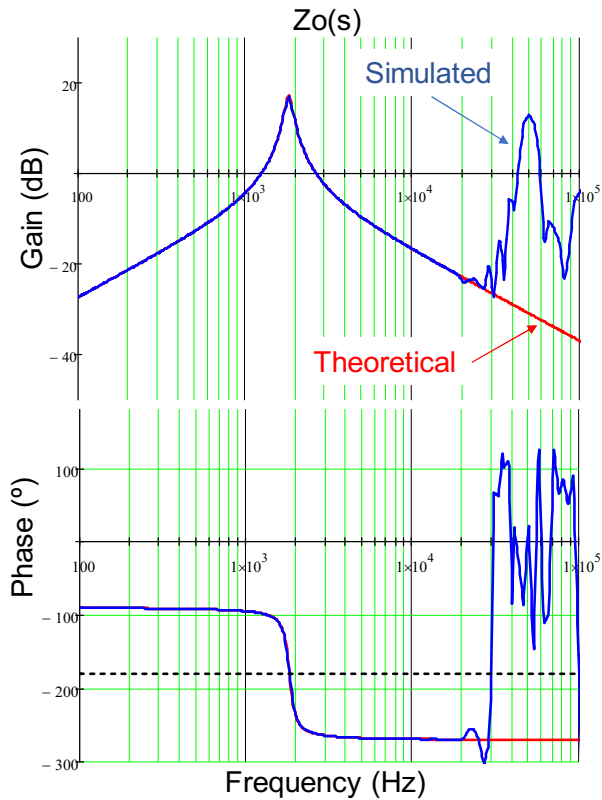

Figure 17. BBMSF converter output impedance, Zo(s)

In Figure 15 - Figure 17 are depicted the small-signal transfer functions corresponding to expressions (32) - (34) respectively. The theoretical representation is depicted in red colour, whereas the simulated one is depicted in blue colour. By comparing both, the theoretical and simulated graphs, it can be noted that the theoretical results fit the simulated ones up to around 20 kHz, which is almost half of the switching frequency. From this frequency on, the simulation results are no longer representative.

As expected from (32), the output voltage-duty cycle frequency response is a second order system with two poles and no zeroes, see Figure 15. This behaviour is the same as the Forward converter. The fact that the BBMSF converter does not have zeroes, especially right half plane (RHP) zeroes, allows to achieve better dynamic performances and simplifies the regulator design.

## 3. EXPERIMENTAL VERIFICATION

### 3.1 CONVERTER DESIGN

#### 3.1.1 The case of study / application environment

A 100kW grid-tied photovoltaic installation with DMPPT architecture is considered as the field of application for the proposed converter, see Figure 2. The central inverter operating voltage is critical to establish the installation configuration, due to it fixes the string voltage. For this study, the *FREESUN LVT FS0100* inverter is selected [36]. The nominal input voltage of this inverter is 600V. Once the central inverter is selected, several PV panels can be chosen. Depending on the type of PV panel, different maximum power point, open circuit and short circuit characteristics must be taken into account. The 225W *SKJ60P6L* PV panel, from Siliken, is the one included in this study [37].

For generating the 100kW with the selected PV panels, 450 units of them are needed. Thanks to the high versatility of the BBMSF converter, which allows both voltage step-up and step-down, a multiple number of strings and a multiple number of PV panels per string configurations are valid. In this case, the selected DMPPT configuration is formed by 25 strings with 18 PV panels in series per string.

A more interesting analysis can be done while considering the mismatching effect between PV panels. Two different scenarios have been defined, depending on the percentage of shaded PV panels. In the Scenario 0, there are no shaded PV panels. In the Scenario 1, a **25%** percentage of the PV panels are affected by shadows. As it is described in [2], a shadow in a PV panel can drastically reduce both, the PV panel voltage and the generated power. A worst-case scenario regarding these two maximum power point characteristics has been considered to obtain the shaded PV panel specifications for the converter design. Every shaded

PV panel has, therefore, approximately, half of the nominal output voltage and a third of the output power [2]. For the sake of simplicity, it is considered that all the PV panels are exactly equals. Two different voltage and power characteristics are therefore defined, one for the non-shaded PV panels (29.3V and 225W) and the other one for the shaded ones (15V and 67.5W).

In a DMPPT architecture, see Figure 2, all the MIC connected to the same string share their output current:

$$i_{string} = \frac{p_{string}}{v_{string}} \tag{35}$$

Being $p_{string}$ and $v_{string}$ the power generated in the string and the voltage of the string respectively. Because central inverter sets the string voltage, a variation in the power generated by a PV panel varies the string current. Considering the ideal case, where the efficiency of the MIC is 100%, the power generated by the PV panel is the same as the power delivered to the string; i.e. $P_{in} = P_{out}$. Taking into account this condition and (35), the output voltage of each converter can be defined by the relationship between the power delivered by the PV panel connected to it and the whole power generated in the string.

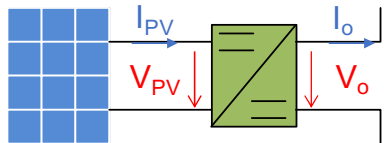

$$V_o = \frac{V_{PV} \cdot I_{PV}}{I_o} = \frac{V_{PV} \cdot I_{PV}}{p_{string}} \cdot v_{string} = \frac{p_{PV}}{p_{string}} \cdot v_{string} \tag{36}$$

Figure 18. Detail of the connection between the PV panel and de MIC

For Scenario E1, applying the power reduction in the shaded PV panels, the delivered power per string, $p_{string}$, can be obtained. Finally, with (36), the output voltage of the converters under both scenarios are obtained.

Table V summarises the case of study specifications.

Table V. Case of study parameters summary

| Converter | Parameter | Scenario E0 | Scenario E1 |
|---|---|---|---|
| Non-shaded PV panel | Power (W) | 225 | 225 |
| | $V_{PV}$ (V) | 29.3 | 29.3 |
| | $V_o$ (V) | 33.3 | 40.404 |
| | $I_{string}$ (A) | 6.75 | 5.569 |
| Shaded PV panel | Power (W) | N/A | 67.5 |
| | $V_{PV}$ (V) | N/A | 15 |
| | $V_o$ (V) | N/A | 12.121 |
| | $I_{string}$ (A) | N/A | 5.569 |

It is noteworthy that, in strings with shaded PV panels, the converters attached to the non-shaded PV panels have to step-up their output voltages whereas the ones attached to the shaded PV panels have to step-down their output voltages. This performance can only be achieved with a voltage step-up and step-down converter.

From the case of study conditions summarised in Table V, the specifications required for the converter design can be defined, see Table VI.

Table VI. Specifications for the converter design

| Parameter | Specification | Parameter | Specification |
|---|---|---|---|
| $V_i$ (V) | [15-29.3] | $D_{max}$ | 0.72 |
| $V_o$ (V) | [12-42.2] ±2% | $P_{omax}$ (W) | 225 |
| $D_{min}$ | 0 | $P_{omin}$ (W) | 60 |

### 3.1.2 Components selection

For the selection of the BBMSF converter components, the electrical stresses in each element have to be obtained. The maximum voltage and current values can be obtained from the expressions described in section 2.2. Table VII shows the values of interest of each element. As previously mentioned in Table IV, the switching frequency is $f_s = 50\ kHz$. Variables without a particular subscript refers to mean values.

Table VII. Electrical stresses in the BBMSF components.

| Component | Voltages (V) | Currents (A) | Other characteristics | Selected Part-reference |
|---|---|---|---|---|
| $C_i$ | $V_{Ci}^{(*)} = 35.16$ | $I_{Ci\_rms} = 7.604$ | $C_{i\_min} = 183.7\mu F$ | C_50SVPF68M *(x4)* |
| $C_o$ | $V_{Co}^{(*)} = 48.48$ | $I_{Co\_rms} = 1.22$ | $C_{o\_min} = 21.91\mu F$ | C_EEHZA1J560P *(x2)* |
| S | $V_{S\_OFF1} = 117.288$<br>$V_{S\_OFF2} = 29.3$ | $I_{S\_peak} = 19.061$<br>$I_{S\_rms} = 10.895$ | $R_{DS\_on} = 9.6m\Omega$<br>$Q_g = 65nC$ | IPB107N20N3 |
| $D_1$ | $V_{D1} = 175.98$ | $I_{D1} = 3.84$<br>$I_{D1\_peak} = 8.864$ | $V_f \approx 1.2V$ | C3D08065E |
| $D_2$ | $V_{D2} = 58.6$ | $I_{D2} = 3.319$<br>$I_{D2\_peak} = 8.864$ | $V_f \approx 0.33V$ | V40D100C-M3/I |
| $D_d$ | $V_{Dd\_ON} = 39.057$<br>$V_{Dd\_OFF2} = 29.3$ | $I_{Dd} = 0.389$<br>$I_{Dd\_peak} = 4.518$ | $V_f \approx 1.2V$<br>$D_2^{(**)} = 0.18$ | C3D06065E |
| L | $V_{L\_ON} = 25.267$<br>$V_{L\_OFF} = 42.2$ | $\Delta I_L = 4.227$<br>$I_L = 6.75$<br>$I_{L\_rms} = 6.859$ | $L = 68\mu H$<br>$DCR = 27.3m\Omega$ | 74435586800 |
| Autotransformer ($L_m$) | $V_{Lm\_ON} = 35$<br>$V_{Lm\_OFF1} = 105$ | $I_{Lm} = 0.696$<br>$I_{Lm\_peak} = 1.504$<br>$I_{Lm\_rms} = 0.866$ | $n = 1$<br>$n_d = \frac{1}{3}$<br>$L_m = 250\mu H$<br>$L_{leakage} = 610nH$<br>$DCR\_pri = 15m\Omega$<br>$DCR\_sec = 17.2m\Omega$<br>$DCR\_ter = 8.5m\Omega$ | Non-standard part by $RENCO^{(***)}$. |

(*) A 20% safety margin is considered in the input and output voltages, see Table VI.

(**) This $D_2$ refers to the autotransformer reset time, see Figure 8. The value shown corresponds with the maximum value, obtained in the Scenario 1.

(***) Manufactured and distributed by RENCO ELECTRONICS, INC., by following our specifications.

Besides the components summarised in Table VII, snubber nets, a driver and a current sensor have been designed and implemented in the BBMSF prototype, see Figure 19.

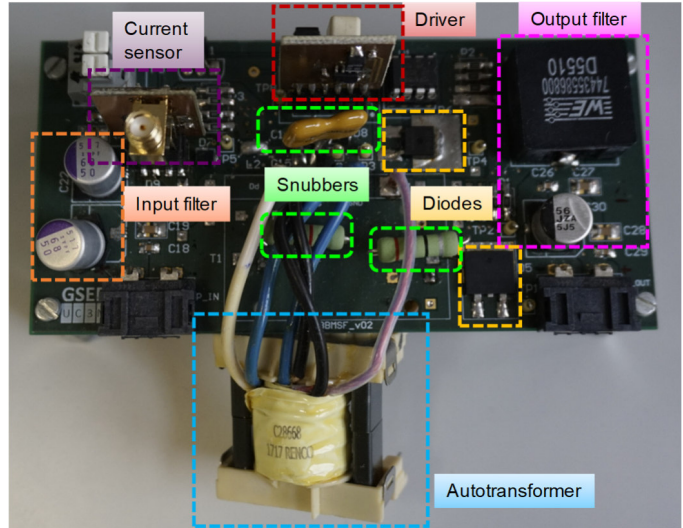

Figure 19. BBMSF prototype.

Typical R-C dissipative snubber net is added in parallel to the $D_2$ diode. For reducing the voltage spike on the MOSFET, both a backwards-regenerative and dissipative snubber nets are used. Figure 20 shows both typical electrical schemes.

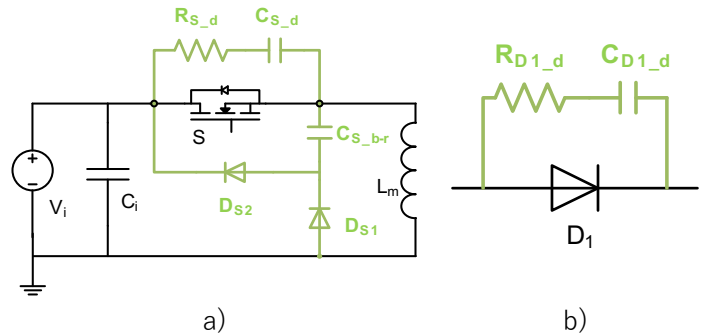

Figure 20. Snubber nets electrical schemes. a) MOSFET snubber nets; b) $D_1$ diode snubber net

The designing process for the snubber nets is well described in [38]. The suffix d refers to the dissipative snubber net type, whereas the b-r suffix refers to the backwards-regenerative one. Table VIII summarises the selected capacitors and resistors employed as snubber nets.

Table VIII. Parameters of the snubber nets added to the BBMSF prototype

| Component | Value |
|---|---|
| $R_{S\_p}$ | 22Ω |
| $C_{S\_p}$ | 2.2nF |
| $C_{S\_b-r}$ | 1nF |
| $R_{D2\_p}$ | 22Ω |
| $C_{D2\_p}$ | 2.2nF |

Due to the MOSFET position, an isolated driver has been designed by using a pulse transformer. The driver electrical scheme and a picture are shown in Figure 21.

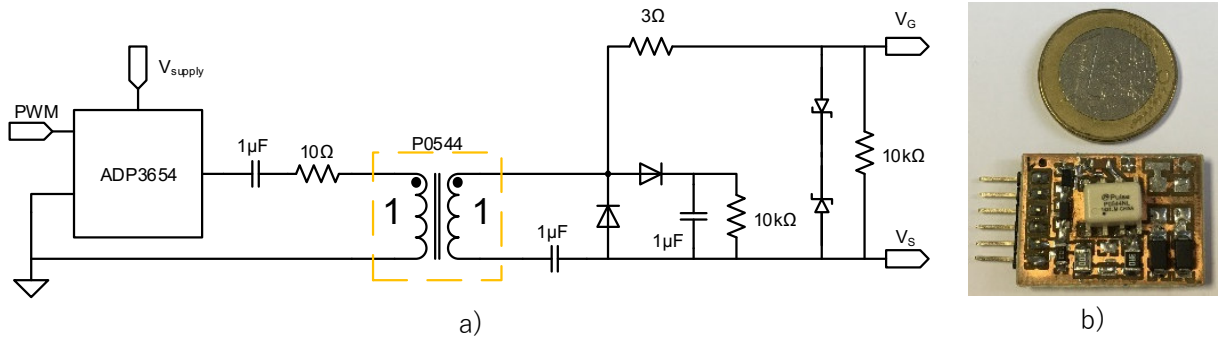

Figure 21. Driver electrical scheme in a); Driver picture of the prototype in b)

The current sensor is based in a current-sensing transformer, see Figure 22.

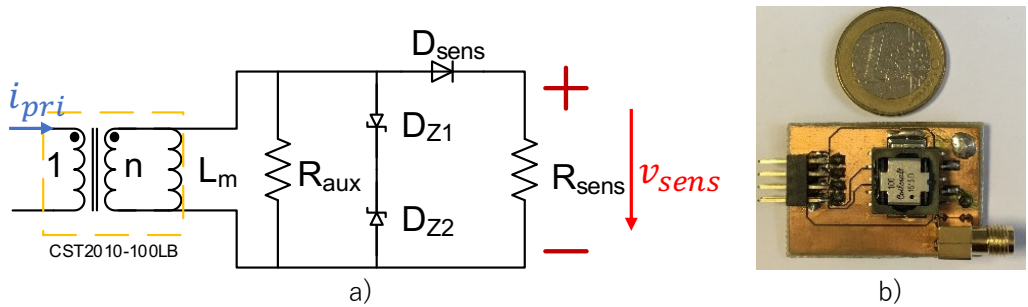

Figure 22. Current sensor electrical scheme in a); current sensor picture of the prototype in b)

Where the turn ratio of the current-sensing transformer is $n = 100$ and the sensing resistance value is $R_{sens} = 50\ \Omega$. Due to these two parameters, the current through the transformer primary side is, in amperes, twice the volts sensed between the sensing resistance terminals (37).

$$i_{pri} = 2 \cdot v_{sens} \tag{37}$$

## 3.2 TIME DOMAIN MEASUREMENTS

This section shows the time domain waveforms of the BBMSF converter MOSFET and diodes. The test conditions are the same as described in the Scenario E0, see Table V. The current waveforms are depicted in pink colour whereas the voltage ones are depicted in yellow colour.
In Figure 23 are shown the S MOSFET current and voltage waveforms.

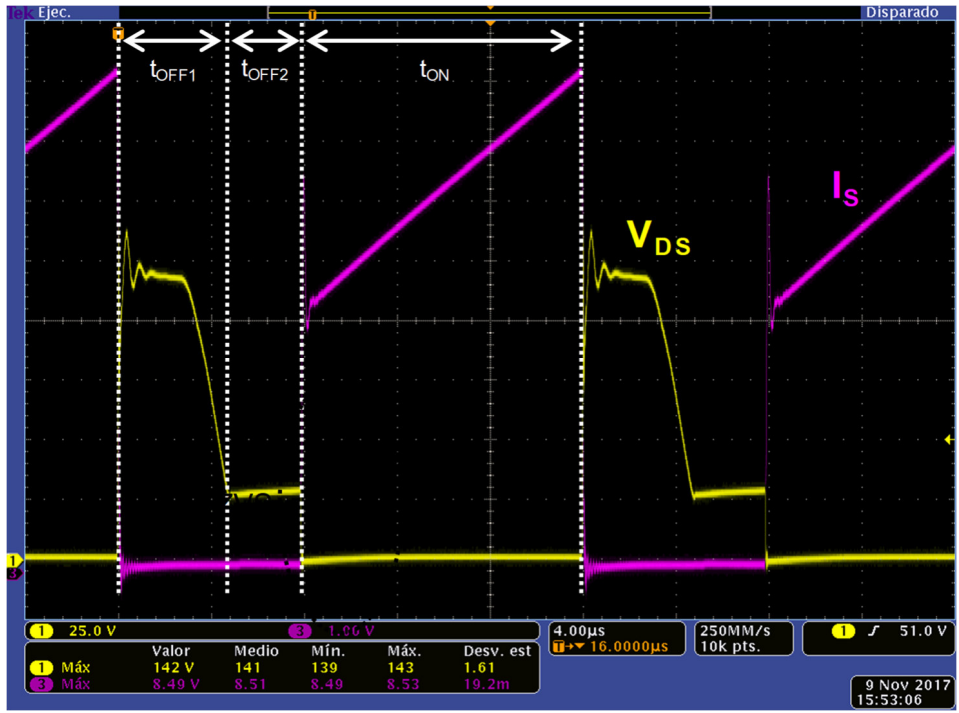

a)

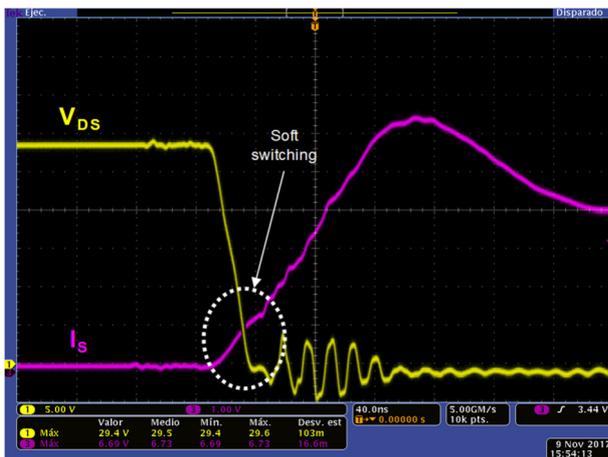

b)

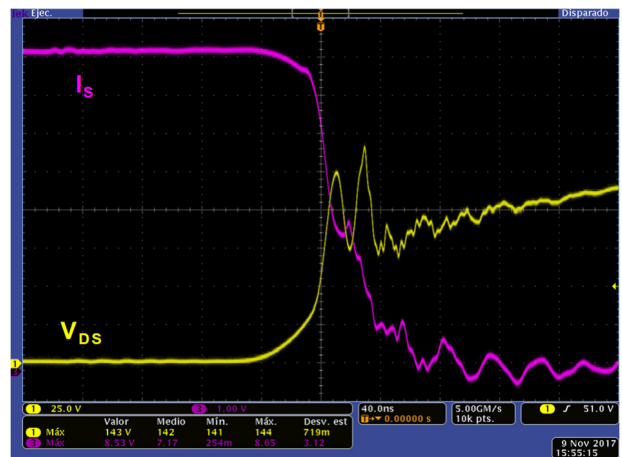

c)

Figure 23. MOSFET waveforms. $V_{DS}$ in yellow colour and $I_S$ in pink colour. a) complete switching period; b) turn-on detail; c) turn-off detail

It is noteworthy that, although snubber nets are included to mitigate the voltage spikes during the switching-off transient, the peak voltage reached is around 20% higher than the theoretical $V_{S\_OFF1}$ voltage, see (25). It is well known that this spike is caused by the autotransformer leakage inductance, through the snubber net and the MOSFET output capacitance ($C_{oss}$). Even higher voltage stress appears on the MOSFET when no snubber nets are used, yielding in worst cases the component breakdown. In addition to the voltage spike in $V_{DS}$, a current spike can also be appreciated in the switching-on transient, see Figure 21a). The energy stored in the snubber and MOSFET capacitors is delivered to the autotransformer abruptly when the MOSFET is turned-on, occasioning the current overshoot. It is important to highlight that this Scenario E0 is the worst scenario regarding the currents flowing through the BBMSF converter components.

As it can be seen in Figure 23.b) the low rising current slope allows the turn-on soft switching in the BBMSF converter. This soft switching characteristic can be considered as almost-Zero Voltage Switching (ZVS).

The leakage inductance is responsible of this "slow" transient. On the other hand, it produces the $V_{DS}$ voltage spike and increases the MOSFET switching losses during the turn-off, see Figure 23.c).

Figure 24 depicts the $D_1$ diode waveforms.

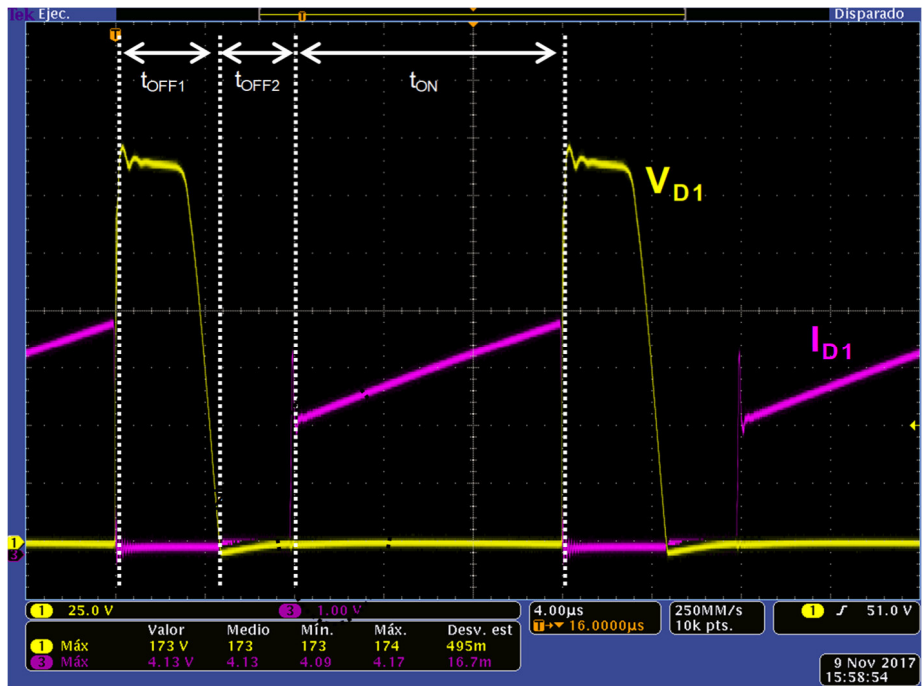

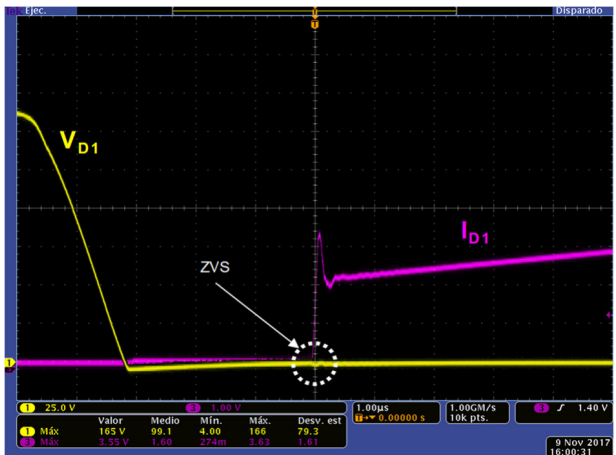

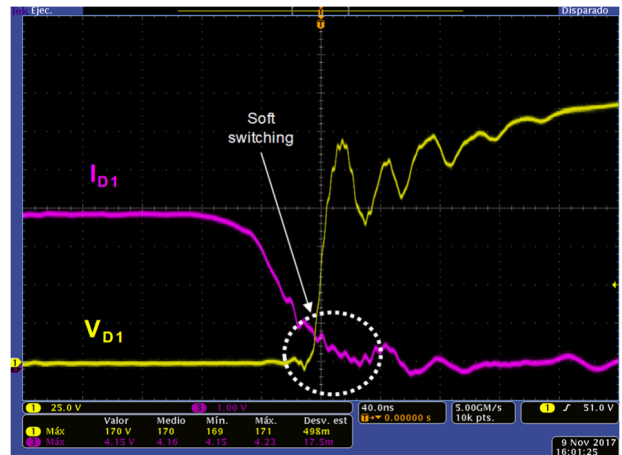

Figure 24. $D_1$ diode waveforms. $V_{D1}$ in yellow colour and $I_{D1}$ in pink colour. a) complete switching period; b) turn-on detail; c) turn-off detail

The same voltage and current spikes effect as in the MOSFET occurs on the $D_1$ diode. It is due to, in fact, the $D_1$ diode is in parallel with the MOSFET during the $t_{OFF}$ interval.

As it can be seen in Figure 24b), this device has ZVS at its switching-on transient. In Figure 24c) almost-Zero Current Switching (ZCS) can be seen during its switching-off. Therefore, very low switching losses are expected in the $D_1$ diode.

Although it is small and therefore hard to appreciate in Figure 24a) and b), there is current flowing through the $D_1$ diode during the $t_{OFF2}$ interval.

Figure 25 depicts the $D_2$ diode waveforms.

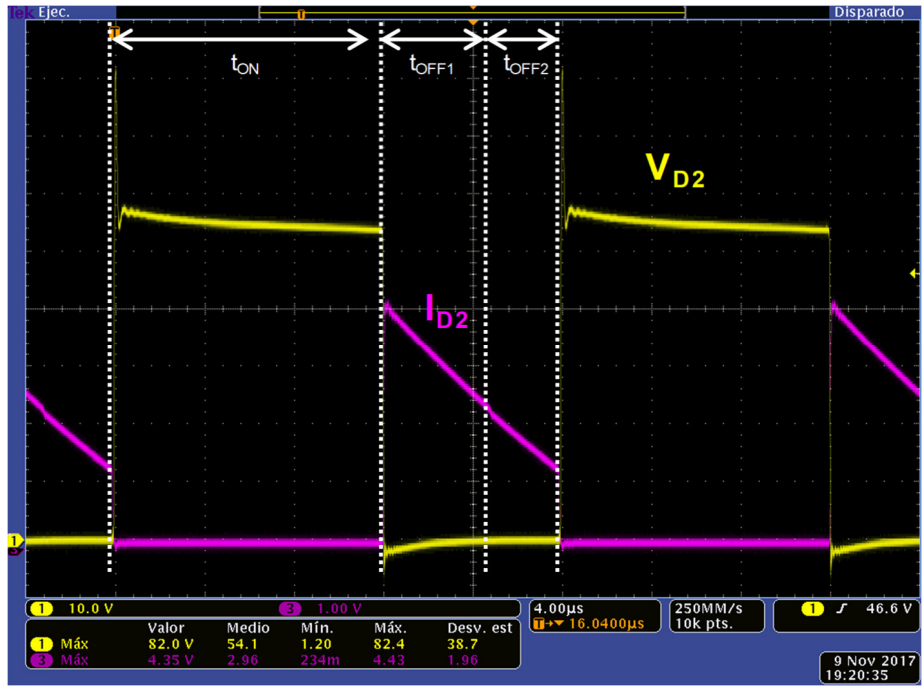

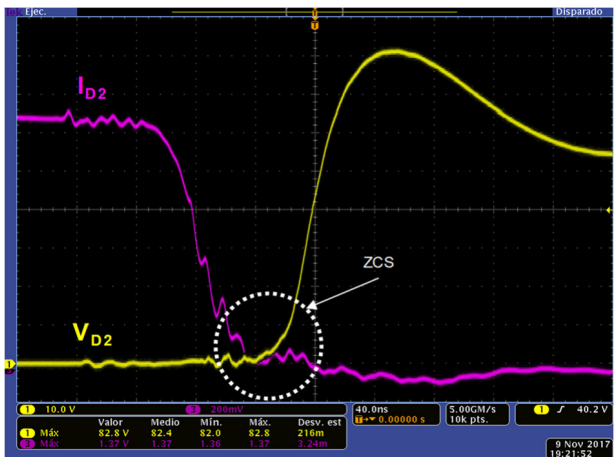
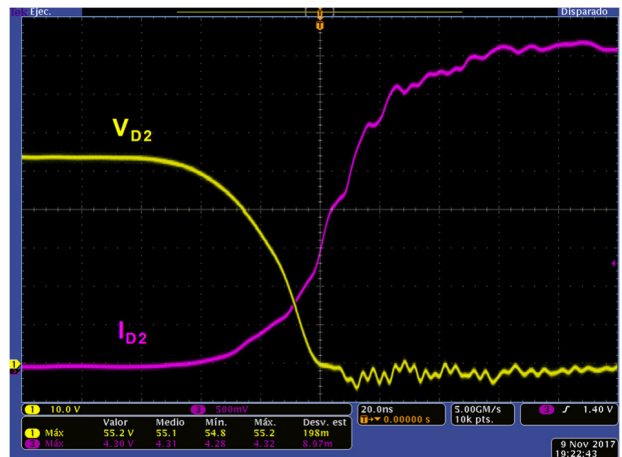

Figure 25. $D_2$ diode waveforms. $V_{D2}$ in yellow colour and $I_{D2}$ in pink colour. a) complete switching period; b) turn-on detail; c) turn-off detail

Although a snubber net is also added in parallel to this diode to reduce the voltage spike, see Table VIII, a voltage overshoot of 40% over the nominal $V_{D2}$ voltage, see Table VII, can be seen in Figure 25a) and b).

It is expected that, as in Figure 10, the $V_{D2}$ voltage remains constant during the $t_{ON}$ time. On the contrary, a slight negative slope can be appreciated in $V_{D2}$, see Figure 25a). This fact is explained by the voltage drop on the MOSFET $R_{DS\_ON}$, that increases as the current $I_S$ does. Therefore, the higher the current through the MOSFET, the higher the voltage drop on the MOSFET and lower the voltage at the autotransformer primary side. This voltage reduction at the autotransformer primary side yields in a reduction in $V_{D2}$.

Low $D_2$ switching losses are expected thanks to the ZCS at the $t_{ON}$ transition, see Figure 25b).

Although the output inductance current waveform is not measured, it can be easily obtained as the $I_{D1}$ and $I_{D2}$ junction.

The last oscilloscope measurements refer to the $D_d$ diode, see Figure 26.

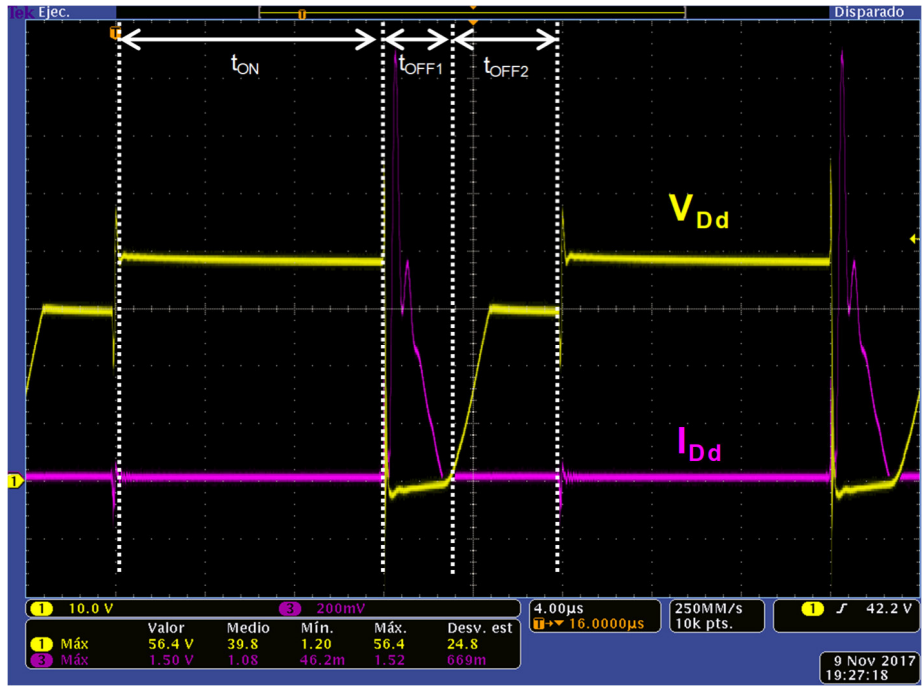

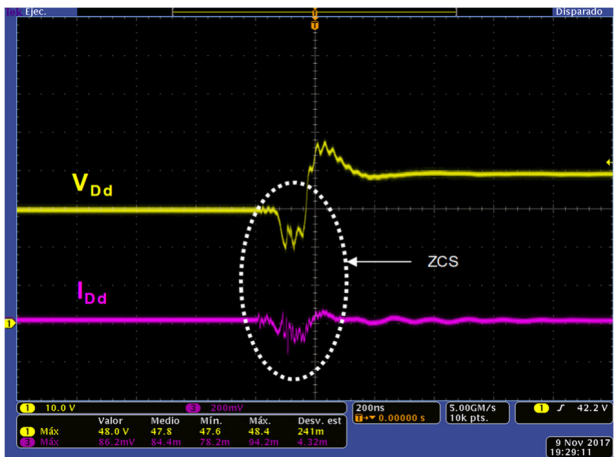

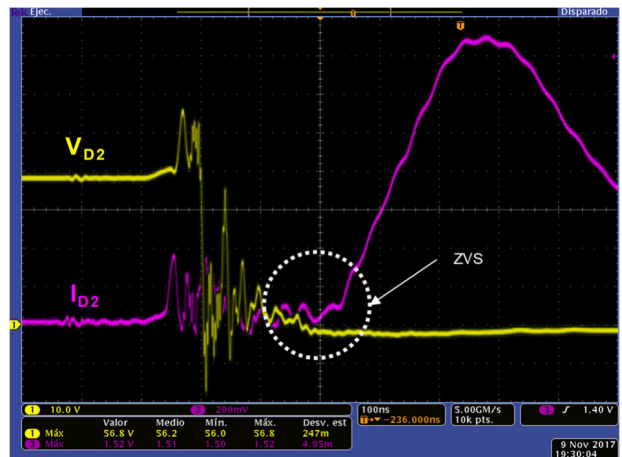

Figure 26. $D_d$ diode waveforms. $V_{Dd}$ in yellow colour and $I_{Dd}$ in pink colour. a) complete switching period; b) turn-on detail; c) turn-off detail

Once the leakage inductance energy is discharged into the MOSFET, $D_1$ and snubber capacitors, the $L_m$ reset begins. As described in section 0, the magnetising inductance energy flows to the input capacitors through the reset winding and $D_d$ diode.

It can be seen in Figure 26.c) that the time delay caused by the leakage inductance reset allows the $D_d$ diode
to switch with ZVS when the MOSFET is switched-off. When the MOSFET is switched-on again, new $t_{ON}$ interval, the Dd diode has ZCS, see Figure 26.b). It is noteworthy that, by Figure 11, at the end of the $t_{OFF1}$ interval, $V_{Dd}$ is different from zero.

Therefore, very low switching losses are expected in the $D_d$ diode.

Comparing the current and voltage waveforms shown in Figure 24 - Figure 26, with the theoretical ones, see Figure 9 - Figure 11 respectively, the steady-state theoretical analysis is verified, as well as the proper operation of the BBMSF converter.

### 3.2.1 Efficiency analysis

A set of measurements is carried out for each one of the output voltage scenarios to obtain the efficiency characteristics of the BBMSF converter, Table V. The efficiency measurements, shown in Figure 27, are obtained using the Yokogawa WT3000. The operation points described in Table V are highlighted in the following Figure 27.

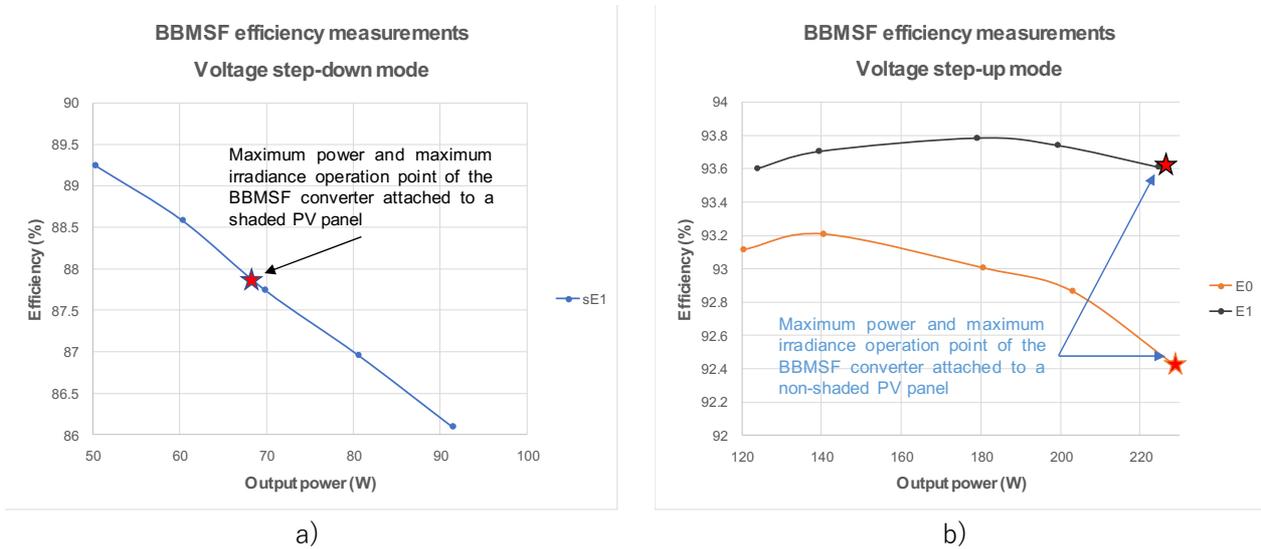

a)                        b)

Figure 27. BBMSF efficiency measurements. A) Efficiencies for the shaded PV panel converter; b) Efficiencies for the non-shaded PV panel converter

In the legend of the graph, the "s" suffix refers to the converters attached to shaded PV panels, see Table V.

As it can be seen in Figure 27.b), for the BBMSF converter voltage step-up mode, the higher the output voltage, the higher the efficiency of the BBMSF converter. It is noteworthy that the converter output voltage in Scenario E1 is higher than in Scenario E0 for not shaded panels. It is due to the conduction losses are more relevant than the switching losses. The BBMSF converter efficiency in step-down mode, i.e. when it is connected to a PV panel in shaded conditions, is the lowest, see Figure 27.a).

A thermal image is taken, see Figure 28, with the purpose of better understanding how the power losses are distributed in the BBMSF converter prototype. The test conditions are the specified for the Scenario E0, and the image is captured after three minutes delivering its maximum power, meaning 225W.

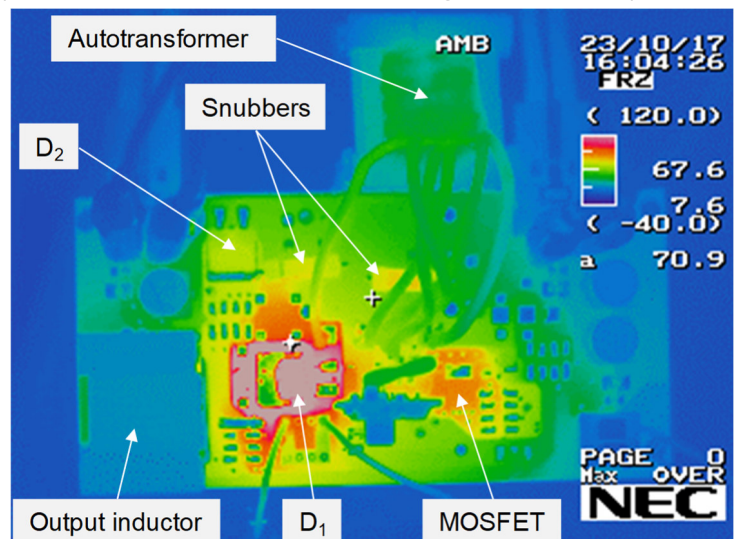

Figure 28. BBMSF converter thermal capture at 225W after three minutes working in the Scenario E0

As it can be seen, the element with a higher temperature is the $D_1$ diode. It means that most of the power losses are focused on this component. It is due to this device has a high forward voltage characteristic, see Table VII, and its losses depend on the average current flowing through it. The selection of another $D_1$ component with lower forward voltage could increase the converter efficiency. In any case, the two characteristics that have been decisive for selecting this device are the high breakdown voltage and the very low capacitance. As long as another diode has these characteristics and lower forward voltage, it should be more convenient than the C3D08065E diode. The snubber resistors are also dissipating a part of the power. The maximum measured efficiency is 93.6%, and it is quite constant from half load to full load, in the case of E0.

The Buck-Boost converter, the Cúk converter with a coupled inductor and the Non-Inverting Buck-Boost converter have been considered as the main MIC competitors, due to all of them are non-isolated voltage step-up and step-down converters, commonly applied to photovoltaic systems. Table IX includes a comparison between them and the BBMSF converter.

Table IX. Comparison between the BBMSF converter with the Buck-Boost converter, the Cúk converter with a coupled inductor and the Non-Inverting Buck-Boost converter

| Parameters | BBMSF | Buck-Boost [30] | Cúk with coupled inductor [30] | Non-Inverting Buck-Boost [27] |
|---|---|---|---|---|
| Magnetic components | 2 | 1 | 1 | 1 |
| Active switches | 1 | 1 | 1 | 4 |
| Drivers | 1 | 1 | 1 | 4 |
| Passive switches | 3 | 1 | 1 | 0 |
| Dynamic performances | No-RHP zero | RHP zero | RHP zero | RHP zero |
| Efficiency (%) | 93.6 | 87.2 | 93.1 | 98.5 |

As it can be deduced from the comparison shown in Table IX, the efficiency of the BBMSF converter is comparable with the efficiencies obtained with other topologies for this application [30]. Although the number of components in the Cúk converter with coupled inductors is lower than in the BBMSF, their diode, main capacitor and active switch suffer also high electrical stresses. Moreover, its dynamic response is slower due to the RHP zero in the small signal transfer functions. Recent works have reported higher efficiencies [14] and [27], but including four controlled switches and drivers. In comparison to them, the semiconductor count of the BBMSF is limited to three diodes and one power MOSFET, resulting in potential cost reductions (it only needs one driver). Besides, the size of the output filter capacitors in the BBMSF converter is lower due to its output filter inductor. The reduced component count, in comparison with the Non-Inverting Buck-Boost converter, can also affect the reliability of the overall system positively. As for the Cúk converter and Buck-Boost converter, the Non-Inverting Buck-Boost converter suffers from RHP zero in their small signal transfer functions, limiting their dynamic performance.

## 4. CONCLUSIONS

A full theoretical analysis of the Buck-Boost Modified Series Forward (BBMSF) converter is carried out in this paper, including the steady-state analysis as well as the small-signal analysis in continuous conduction mode.

The time domain analysis covers the output-input voltage transfer function, the principle of operation and all the voltage and current expressions in each component of the BBMSF converter. In addition, a power analysis shows the magnetically processed output power ratio.

In the frequency domain analysis, the output voltage – duty cycle ($G_{vd}(s)$), the audiosusceptibility ($G_{vv}(s)$) and the output impedance ($Z_o(s)$) small signal transfer functions are obtained and validated through simulation results. The dynamic performances of the BBMSF converter are similar to the well-known Forward converter ones. Therefore, good dynamic performances can be achieved with the BBMSF converter.

A 225W prototype has been built according to the specifications of a 100kW PV grid-tied installation. The case of study considers the effect of a mismatching caused by shadows. Two different scenarios have been defined with different shaded ratios. The measured results verify the previous analysis. The soft switching characteristics are highlighted for each diode. The effect of some parasitic inductances, not considered in the theoretical analysis, are shown and explained with the measured waveforms. Besides, a set of measurements are carried out to plot the BBMSF converter efficiency characteristics. The highest measured efficiency is 93.6%. A thermal picture reveals that most of the power losses are in the $D_1$ diode, due to its high forward voltage characteristic.

Summarizing, the main advantages of the BBMSF converter are its step-up and step-down voltage transfer function; the ZVS, ZCS and soft switching characteristics in the diodes and MOSFET, yielding in efficiencies up to 93.6%; the use of an autotransformer, with better performances than a typical Forward transformer; and the good dynamic performances, like the Forward converter ones. The use of a single controlled switch can result in a potential reduction of cost and increase of reliability. On the other hand, the main drawbacks of the introduced converter are the high voltage stresses in the $D_1$ diode and MOSFET. Devices able to withstand higher voltages but with worse performances are selected. Also, snubber networks must be added to mitigate these voltage spikes, yielding in an efficiency reduction. Although it is not a requirement for the field of application, the output of the converter is not isolated from the input, due to the use of an autotransformer.

## Acknowledgements

This work has been supported by the Ministry of Economy and Competitiveness and FEDER funds through the research project "Storage and Energy Management for Hybrid Electric Vehicles based on Fuel Cell, Battery and Supercapacitors" - ELECTRICAR-AG- (DPI2014-53685-C2-1-R).